\documentclass{elsart}

\usepackage{natbib}
\usepackage{amssymb}
\usepackage{graphicx}

\newcommand{\ie}{i.e.\/}
\newcommand{\eg}{e.g.\/}
\newcommand{\keyw}[1]{{\bf #1}}

\begin{document}

\begin{frontmatter}

\title{LUDWIG: A parallel Lattice-Boltzmann code for complex fluids}

\author[EPCC]{Jean-Christophe Desplat\thanksref{corres}}
\author[DPA]{Ignacio Pagonabarraga}
\author[DPA]{Peter Bladon}

\address[EPCC]{Edinburgh Parallel Computing Centre, King's Buildings,
	the University of Edinburgh, Edinburgh EH9 3JZ, UK}
\address[DPA]{Department of Physics and Astronomy, King's Buildings,
	the University of Edinburgh, Edinburgh EH9 3JZ, UK}
\thanks[corres]{author for correspondence (E-mail: j-c.desplat@epcc.ed.ac.uk;\\
	Tel.: +44 131 650 6716; Fax: +44 131 650 6555)}

\begin{abstract}
This paper describes {\em Ludwig}, a versatile code for the simulation
of Lattice-Boltzmann (LB) models in 3-D on cubic lattices. In fact
{\em Ludwig} is not a single code, but a set of codes that share
certain common routines, such as I/O and communications. If {\em
Ludwig} is used as intended, a variety of complex fluid models with
different equilibrium free energies are simple to code, so that the
user may concentrate on the physics of the problem, rather than on
parallel computing issues. Thus far, {\em Ludwig}'s main application
has been to symmetric binary fluid mixtures.  We first explain the
philosophy and structure of {\em Ludwig} which is argued to be a very
effective way of developing large codes for academic consortia.  Next
we elaborate on some parallel implementation issues such as parallel
I/O, and the use of MPI to achieve full portability and good
efficiency on both MPP and SMP systems.  Finally, we describe how to
implement generic solid boundaries, and look in detail at the
particular case of a symmetric binary fluid mixture near a solid
wall. We present a novel scheme for the thermodynamically consistent
simulation of wetting phenomena, in the presence of static and moving
solid boundaries, and check its performance.
\end{abstract}

\end{frontmatter}

\section{Objectives}
\label{sec:objectives}
The objective of the work described here has been to develop a general
purpose parallel Lattice-Boltzmann code (LB), called {\em Ludwig},
capable of simulating the hydrodynamics of complex fluids in 3-D. Such
a simulation program should eventually be able to handle
multicomponent fluids, amphiphilic systems, and flow in porous media
as well as colloidal particles and polymers. In due course we would
like to address a wide variety of these problems including detergency,
binary fluids in porous media, mesophase formation in amphiphiles,
colloidal suspensions, and liquid crystal flows. So far, however, we
have restricted our attention to simple binary fluids, and it is this
version of the code that will be described below in more
detail. Nonetheless, the generic elements related to the structure of
the code are valid for any multicomponent fluid mixture, as defined
through an appropriate free energy, expressed as a functional of fluid
density and one or more composition variables (scalar order
parameters). We discuss in some detail also how to include solid
objects, such as static and moving walls and/or freely suspended
colloids, in contact with a binary fluid. More generally, the modular
structure of {\em Ludwig} facilitates its extension to many other of
the above problems without extensive redesign. But note that, with
several of these problems (such as liquid crystal flows which require
tensor order parameters), it is not yet clear how to proceed even at
the serial level, and only first attempts have begun to appear in the
literature~\cite{Julia_prep}.

\section{Lattice Boltzmann model}
\label{sec:model}
The Lattice-Boltzmann model (LB) simulates the Boltzmann equation with
linearized collisions on a lattice~\cite{Higuera}. Both the changes in
position and velocity are discretized.  It can be shown that, at
sufficiently large length and time scales, LB simulates the dynamics
of nearly incompressible viscous flows~\cite{Qian92,Ladd94}. For the
simplest case of a one-component fluid, it describes the evolution of
a discrete set of particle densities on the sites (or {\em nodes}) of
a lattice:
\begin{equation}
\label{eqn:lb}
f_i(\vec{r} + \vec{ c}_i, t + 1) - f_i(\vec{r}, t) =
-\omega \left( f^{\mathrm{eq}}_i(\vec{r},t) - f_i(\vec{r},t) \right)
\end{equation}
The quantity $f_i(\vec{r},t)$ is the density of particles with
velocity $\vec{c}_i$ resident at node $\vec{r}$ at time $t$. This
particle density will, in unit time increment, be convected (or {\em
propagate}) to a neighboring site $\vec{r} + \vec{c}_i$. Here
$\vec{c}_i$ is a lattice vector, or {\em link} vector, and the model
is characterized by a finite set of velocities $\{\vec{c}_i\}$. The
quantity $f^{\mathrm{eq}}_i(\vec{r},t)$ is the \lq equilibrium
distribution' of $f_i(\vec{r},t)$, and is one of the key ingredients
of the model. It characterizes the type of fluid that {\em Ludwig}
will simulate, and determines the equilibrium properties of such a
fluid (see section \ref{ssec:binaries} below). The right hand side of
equation~\ref{eqn:lb} describes a mixing of the different particle
densities, or {\em collision}: the $f_i$ distribution relaxes towards
$f_i^{\mathrm{eq}}$ at a rate determined by $\omega$, the relaxation
parameter. The relaxation parameter is related (through $\eta =
(2\omega^{-1} - 1)/6$) to the viscosity $\eta$ of the fluid, and gives
us control of its dynamics.

To specify a particular model, besides the equilibrium properties
given through $f^{\mathrm{eq}}$, one has to choose the geometry of the
lattice in which the density of particles move. Such a geometry should
specify both the arrangement of nodes and the set of allowed
velocities. The only restrictions in such a choice lie on the fact
that they should have sufficient symmetry to ensure that at the
hydrodynamic level the behavior is isotropic and independent of the
underlying lattice~\cite{Wolfram86}. The hydrodynamic quantities, such
as the local density, $\rho$, momentum, $\rho {\bf v}$ and stress,
$\mathcal{P}_{\alpha\beta}$ are given as moments of the densities of
particles $f_i(\vec{r},t)$~\cite{Qian92,Ladd94}, namely $\sum_i f_i =
\rho$, $\sum_i f_i {\vec c_i} = \rho {\vec v}$, and $\sum _i f_i
c_{i\alpha} c_{i\beta} = \mathcal{P}_{\alpha\beta}$.

The dynamics of LB, as expressed in equation~\ref{eqn:lb}, provides
immediate insight into the implementation and underlying optimization
issues. It is characterized by two basic dynamic stages:
\begin{itemize}
\item the propagation stage (left-hand side of equation~\ref{eqn:lb}),
consisting of a set of nested loops performing memory-to-memory
copies;
\item the collision stage (right hand side), which has a strong degree
of spatial locality and relies on basic add/multiply operations: its
implementation is straightforward and can be highly optimized.
\end{itemize}

\subsection{Binary fluid mixtures}
\label{ssec:binaries}
\begin{figure}[h]
\begin{center}
\includegraphics[scale=0.3]{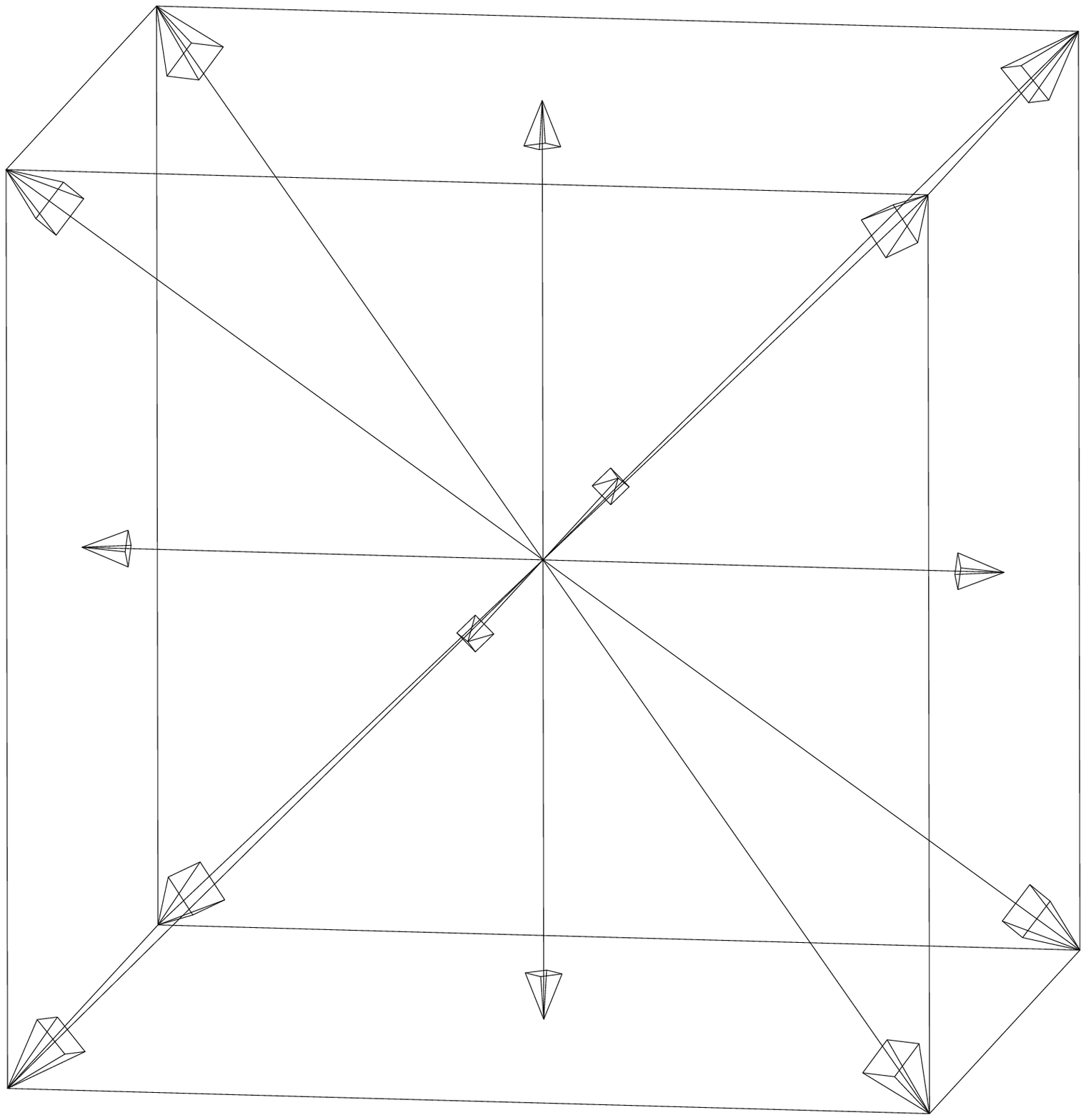}
\includegraphics[scale=0.3]{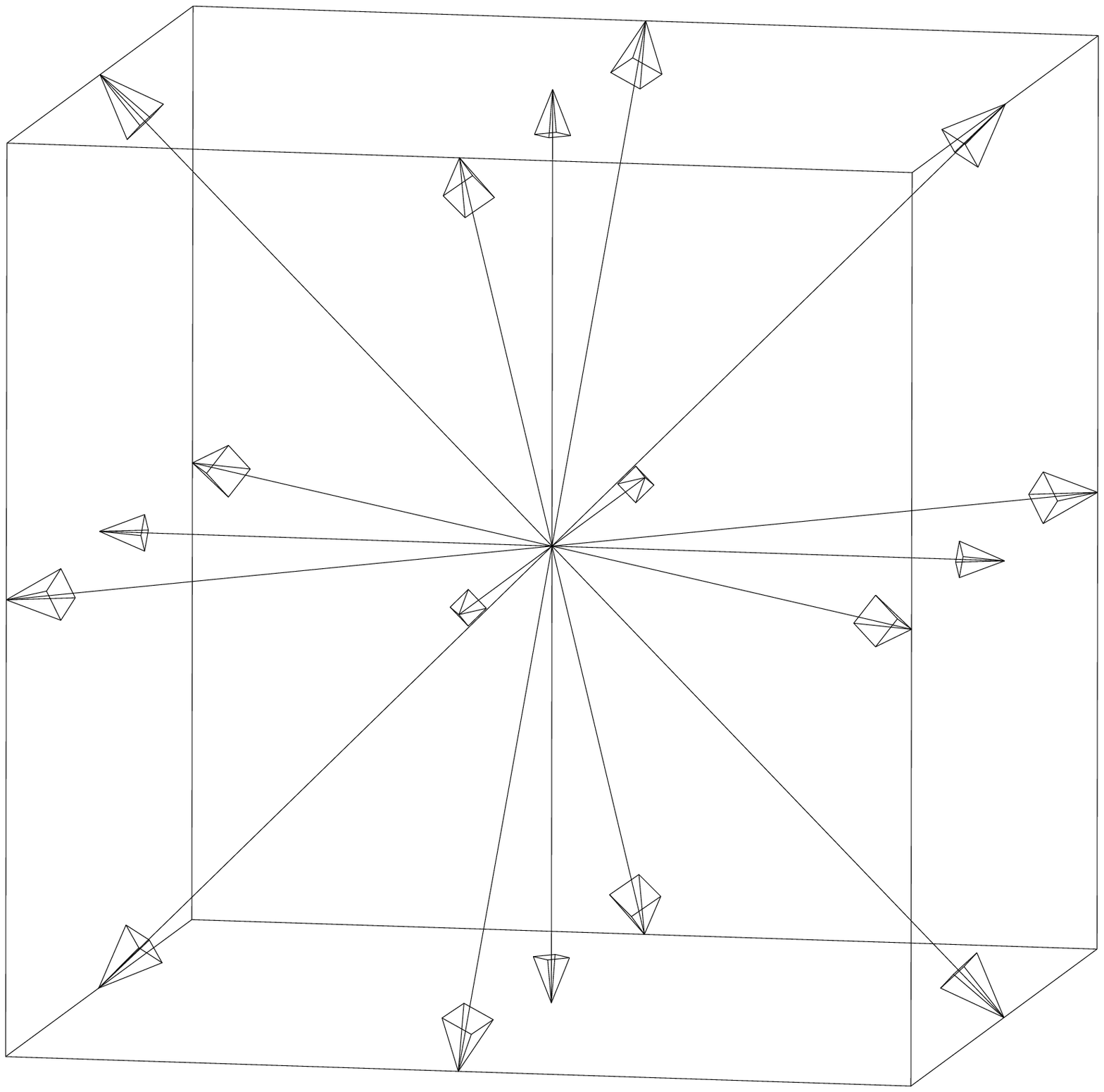}
\end{center}
\caption{D3Q15 (left) and D3Q19 (right) models. The D3Q15 model has
fifteen velocities: one with speed zero (a rest particle), six with
$(speed)^2 = 1$ (to nearest neighbors), and eight with $(speed)^2=3$
(to next next nearest neighbors). The D3Q19 model has nineteen
velocities: one with speed zero (a rest particle), six with $(speed)^2
= 1$ (to nearest neighbors), and 12 with $(speed)^2=2$ (to next
nearest neighbors).}
\label{fig:d3qx}
\end{figure}

The LB model described so far can be extended to describe a binary
mixture of fluids, of tunable miscibility, by adding a second
distribution function, $g_i$\cite{Swift}. (Further distribution
functions would allow still more complicated mixtures to be
described.)  As in single-fluid LB, the relevant hydrodynamic
variables related to the order parameter are also moments of the
additional distribution function $g_i$, namely the composition (order
parameter) $\phi=\sum_i g_i$, and the flux $\sum_i g_i {\vec c_i} =
\phi {\vec v}$.
For each site (including solid sites), the distributions $f_i$ and
$g_i$ are stored in a structure element of type \keyw{Site}:
\begin{tabbing}
xxx \= xxx \= \kill
typedef struct\{\\
   \>Float f[NVEL],\\
   \>      \> g[NVEL];\\
\} Site;
\end{tabbing}
\noindent where \keyw{NVEL} is the number of velocity vectors used by
the model. For example, for the cubic lattices described later on,
where {\em Ludwig} has been implemented so far, the number of velocity
vectors has been 15 and 19. Figure~\ref{fig:d3qx} shows the sets of
velocities for the two 3-D models developed.

We follow the procedure of Swift et al.\/~\cite{Swift} (see
also~\cite{Grubert} for a schematic description) in which $f_i$
describes the density field $\rho$, whilst $g_i$ describes the order
parameter field, $\phi$. Both distribution functions have relaxational
dynamics of the type of equation~\ref{eqn:lb} but are characterized by
different relaxation parameters $\omega_{\rho,\phi}$. The second
relaxation parameter, associated with the order parameter field, will
determine its diffusivity. By studying appropriate moments of the
distribution functions, one can construct a relaxational dynamics that
will describe, in the continuum limit, the dynamics of a
near-incompressible, isothermal binary fluid with an arbitrary local
free energy functional $F[\phi]$.  The model chosen is a symmetric \lq
$\phi^4$' or Cahn-Hilliard type free energy:
\begin{equation}
\label{eqn:fe}
F = \int \mathrm{d}{\mathbf r}\left\{\frac{A}{2}\phi^2+\frac{B}{4}\phi^4
+\frac{\kappa}{2}|\nabla\phi|^2\right\}
\end{equation}
\noindent where $A$, $B$ and $\kappa$ are model parameters. For the
density $\rho$, LB dynamics ensures an ideal gas equation of state,
with a speed of sound equal to $1/\sqrt{3}$ . In practice $\rho$
remains almost constant at a value which we choose to be unity. This
can be done by ensuring that under all conditions the fluid velocity
remains small compared to unity in lattice units. (More generally one
requires velocities small compared to the sound speed.)  For negative
$A$, the above model has two coexisting fluid phases with order
parameter values $\pm \phi^*$; for many problems it is convenient to
set $B = -A$ so that $\phi^* = 1$.

Note also that, although there is a long history of studying $\phi^4$
theory on the lattice, one needs to be aware of possible lattice
artifacts in the thermodynamic, as well as the hydrodynamic, sectors
of the model~\cite{Kendon00}. For example, the coefficient $\kappa$,
determines the thickness $\xi_0 = (\kappa/2|A|)^{1/2}$ and the
interfacial tension $\sigma = (8\kappa |A|^3/9 B^2)^{1/2}$ of the
interface between two fluids~\cite{Swift}. But the thickness must be
kept large enough to avoid a strong anisotropy of the interfacial
tension caused by the underlying lattice. Moreover, the values of the
different parameters should be carefully selected to give the required
compromise between numerical stability and accuracy, and computational
speed~\cite{Kendon00}. However, since the same physical parameters (in
a binary fluid, viscosity, density, interfacial tension) can be
achieved with more than one set of simulation parameters, it is
normally possible to steer around any problems, though they do present
traps for the unwary.  The specific role played by order parameter
mobility in the simulations of binary fluids is discussed
in~\cite{Kendon00,Kendon99}.

We emphasize that {\em Ludwig} is structured so that the free energy
functional can be chosen at will. This is a desirable feature of LB
over, for example, the dissipative particle dynamics algorithm (DPD),
where the free energy being modeled has to be deduced {\em a
posteriori} from the simulation results~\cite{Groot}, although first
attempts are being carried out to allow for free energy {\it a priori}
determination~\cite{Daan}.  The user of the code has to evaluate, from
the free energy and according to a well-established
procedure~\cite{Swift,Kendon00}, the equilibrium distribution
functions $f_i^{\mathrm{eq}}$, $g_i^{\mathrm{eq}}$ for use in the
relaxation equation~\ref{eqn:lb} and the corresponding one for $g_i$.
This data is entered into the subroutine for the collision step. For
example, in the case of the binary fluid mixture in the D3Q15 geometry
one has~\cite{Kendon00}
\begin{equation}
f_i^{\mathrm{eq}} = \rho \omega_{\nu}\left\{A_{\nu} + 3v_{\alpha} c_{i\alpha}
        + \frac{9}{2}v_{\alpha} v_{\beta}c_{i\alpha} c_{i\beta}
        - \frac{3}{2}v^2 + G_{\alpha\beta}c_{i\alpha} c_{i\beta}\right\}.
\end{equation}
Here, ${\nu}$ is an index that denotes the speed, $|{\vec c}_i| = 0,
1, 3^{1/2}$, and $\omega_{\nu}$, $A_{\nu}$ and $G_{\alpha\beta}$ are
constants given by
\begin{equation}
\omega_0 = 2/9;\;\;\; \omega_1 = 1/9;\;\;\; \omega_3 = 1/72,
\end{equation}
\begin{equation}
A_0 = \frac{9}{2} - \frac{7}{2} \mathrm{Tr}\mathcal{P}^{\mathrm{th}};\;\;\;
A_1 = A_3 = \frac{1}{\rho} \mathrm{Tr}
\mathcal{P}^{\mathrm{th}},
\end{equation}
\begin{equation}
G_{\alpha\beta} = \frac{9}{2\rho} \mathcal{P}^{\mathrm{th}}_{\alpha\beta} - 
\frac{3\delta_{\alpha\beta}}{2\rho}\mathcal{P}^{\mathrm{th}}.
\end{equation}
Here $\mathcal{P}^{\mathrm{th}}_{\alpha\beta}$ is the thermodynamic
contribution to the pressure tensor which can be evaluated directly
from the chosen form of the free energy functional~\cite{Swift}, and
for equation~\ref{eqn:fe} it reads
\begin{equation}
\label{eq:ld_pt}
\mathcal{P}^{\mathrm{th}}_{\alpha\beta} = \left\{\frac{\rho}{3} + \frac{A}{2}\phi^2 + \frac{3B}{4}\phi^4
- \kappa\phi\nabla^2\phi - \frac{\kappa}{2}(\nabla\phi)^2\right\}\delta_{\alpha\beta}
+ \kappa(\partial_{\alpha}\phi)(\partial_{\beta}\phi).
\end{equation}
The equilibrium distribution for the order parameter,
$g_i^{\mathrm{eq}}$, is the same as for $f_i^{\mathrm{eq}}$, with
$\mathcal{P}^{\mathrm{th}}_{\mu\nu}$ replaced by $\tilde
M\mu\,\delta_{\mu\nu}$ in the above equations; here $\tilde
M(\omega_\phi^{-1} - 1/2) = M$, where $M$ is the order parameter
mobility~\cite{Kendon00}.

\subsection{Gradient discretization}
For evaluation of $\mathcal{P}^{\mathrm{th}}_{\alpha\beta}$ and other
quantities, we need to compute spatial gradients of $\phi$. To
minimize thermodynamic lattice anisotropies, this is done using a
larger set of links than used for the propagation step; for example on
the D3Q15 lattice we use all 26 (first, second and third) nearest
neighbors so that numerically
\begin{equation}
\label{eq:grad_def}
\partial_{\alpha}\phi(\mathbf{x}) =
        \frac{\sum_i c_{i\alpha}\phi(\mathbf{r}+\mathbf{c}_i)}
        {\sum_i c_{i\alpha}c_{i\alpha}}
\end{equation}
\begin{equation}
\label{eq:lap_def}
\nabla^2\phi(\mathbf{r}) = \frac{1}{9}\left[
        \left(\sum_{i=1}^{26}\phi(\mathbf{r}+\mathbf{c}_i)\right)
        - 26\phi(\mathbf{r}) \right]
\end{equation}
\noindent with an enlarged set $\{{\vec c}_i\}$. Note that these are
not the only possible choices. There may be considerable scope for
further improvement by optimizing the choices made for gradient
discretization, but we leave this for future work.

\subsection{Numerical stability}
One drawback of LB is that, unlike DPD and some other competing
mesoscale techniques, it is not unconditionally stable. However, our
experience suggests that even if the model becomes eventually unstable
for any given set of parameter values, the problem arises so suddenly
that such an instability does not impede collection of robust and
reliable data over long periods
beforehand~\cite{Kendon00}. Nonetheless, it would be very desirable to
have a fully stable version of the algorithm and this might
considerably reduce the time spent in parameter steering exercises
that are currently needed prior to allocating resources for production
runs.

\section{Implementation details}
\subsection{Users' requirements}
\label{ssec:requirements}
The scientific objectives presented in section~\ref{sec:objectives}
are ambitious and imply the availability of a variety of different
features, consistent with a development time stretching across a few
years.  Because none of the LB codes available at the start of the
project had the required features, {\em Ludwig}'s creators decided to
design a new package from scratch whose main characteristic would be
its versatility: {\em Ludwig} had to be capable of producing data of
scientific interest at an early stage of its development, include
built-in support for multiple models and free energies, and be
sufficiently user-friendly to be customisable and usable by
non-programmers.

Great care has thus been taken to use a design which would fulfill all
of the above requirements. The best approach would thus maximize code
re-use to cut down development time, and be portable and extendible to
increase the package's overall lifetime. The portability issues led us
to choose ANSI-C and MPI-1.1. This combination provided the required
features without sacrificing too much in the area of performance. In
addition to portability, MPI also provided valuable features such as
support for user-defined operators (\eg\ to perform global operations
on distributed sets of vectors and tensors) and a high level of
abstraction through the use of derived datatypes. The latter is
particularly valuable to make all the I/O and communication routines
non model-specific.

\subsection{Decomposition strategy}
Due to the magnitude of the system size required to study the
phenomena described in section~\ref{sec:objectives}, it became obvious
from the early stage of the design that {\em Ludwig} would have to be
parallelized in order to provide the required scalability.
Fortunately, the symmetry of the underlying cubic lattice guarantees a
uniform data distribution and hence an equal amount of computations
per lattice site. Indeed, the collision and propagation stages will
take place over all lattice sites, which restricts possible causes of
load imbalance to the introduction of solids objects non-uniformly
distributed across the simulation box.  This pseudo-uniform
distribution of the computations added to the intrinsic locality of
the LB algorithm made {\em Regular Domain Decomposition} the most
suitable decomposition strategy~\cite{Foster95}. In this approach, the
data is geometrically decomposed in equal volumes, which are then
distributed to each processing element (PE).

Obviously the `ideal' load-balancing can only be achieved for a
restricted class of studies such as simulations of spinodal
decomposition. As soon as Lees-Edwards walls or solid objects are
introduced ---whether these are truly moving or freely suspended
particles such as colloids, pseudo-moving plates to shear the system,
or static structures such as a porous network---, the behavior of the
code will be affected as its overall performance becomes restricted by
that of the slowest process. However, most applications described in
section~\ref{sec:objectives} exhibit sufficient symmetry to circumvent
the load imbalance induced by the solid objects. Should the solid
objects be distributed inhomogeneously across the system, it will be
necessary to specify a more adequate processor topology to ensure an
efficient parallelization.

\subsection{Structure}
{\em Ludwig} has been developed with a modular and hierarchical
structure in mind. The current version of the package is composed of
258 functions (over 25,000 lines of code) split in three main
components, as illustrated in figure~\ref{fig:structure}:

\begin{figure}[h]
\begin{center}
\includegraphics{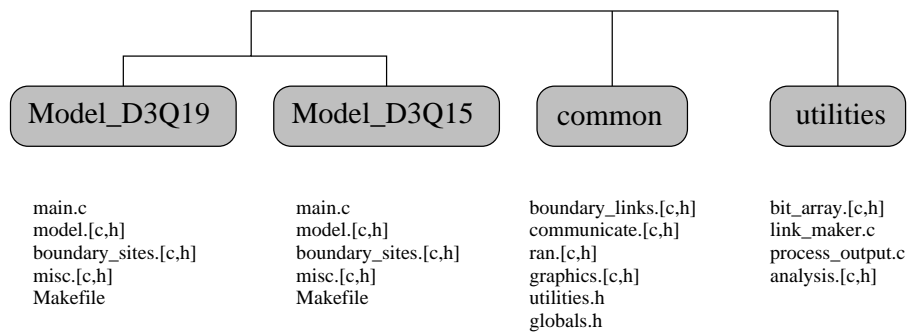}
\end{center}
\caption{Program structure.}
\label{fig:structure}
\end{figure}

\begin{enumerate}
\item Model subdirectories: contain all the model-specific functions
as well as main.c. These model-specific options contain three main
ingredients of the code: the geometry of the lattice, the free energy
that defines the type of fluid to be modeled, and the type of boundary
conditions which determines the interaction with solid objects. Users
can plug-in their own routines in misc.c. Once the model is defined,
the {\em only} modification required to run simulations is to edit
main.c to call the relevant measuring functions.  At this time, the
models available include the lattice geometries D3Q15, and D3Q19 (see
figure~\ref{fig:d3qx} and reference~\cite{Qian92,Ladd94} for their
definitions), although this package has been designed to support
models with subset of velocities beyond the third nearest neighbors.

\item Common subdirectory: contains all the low-level calls such as
the communication layer and the parallel I/O, as well as a set of
routines to provide real-time graphics during simulations. This
functionality proves invaluable to gain a better understanding of the
dynamics and for debugging purposes. These generic functions can be
called by all models.

\item Utilities subdirectory: contains stand-alone pre- and
post-processors for setting-up initial configurations and analyzing
the simulation data.
\end{enumerate}

The non-specificity of the common routines has been made possible by
MPI's high-level features such as derived datatypes, communicators,
and user-defined operators. Indeed, these routines can be used to
define series of generic and opaque objects and operators which can be
accessed and manipulated without the need to know their model-specific
characteristics. In effect, these routines implement a generic,
generalized model, which in some ways is reminiscent of the concept of
objects and methods as implemented in the object-oriented
paradigm. Since the use of object-oriented programming language such
as C++ or java had to be discounted on the ground of their lack of
bindings for MPI-1.1 and their poor performance on HPC systems, the
approach described above is in the authors' opinion the best
alternative.

The main advantage of this modular approach is the fact that the
computational complexity is hidden, which allows the users to
concentrate on the physical analysis of a given system rather than on
implementation issues. Other advantages include code re-use, package
extendibility, portability and efficiency. {\em Ludwig} achieves a
high level of portability: indeed, it has been successfully installed
on a variety of serial and parallel platforms (Cray T3E, T3D and J90,
SGI Origin 2000, Hitachi SR-2201, Sun E-3000/HPC-3500, DEC and Sun
workstations as well as a Linux PC) with no modification required.

\subsection{Solid objects}
\label{ssec:solidobj}
Three different types of solid objects may be defined in {\em Ludwig}:
\begin{enumerate}
\item static solid objects, \eg\ static walls and porous networks;
\item (pseudo-)moving walls, \eg\ to shear the system;
\item (truly-)moving particles, \eg\ freely suspended colloids.
\end{enumerate}

\begin{figure}[h]
\begin{center}
\includegraphics{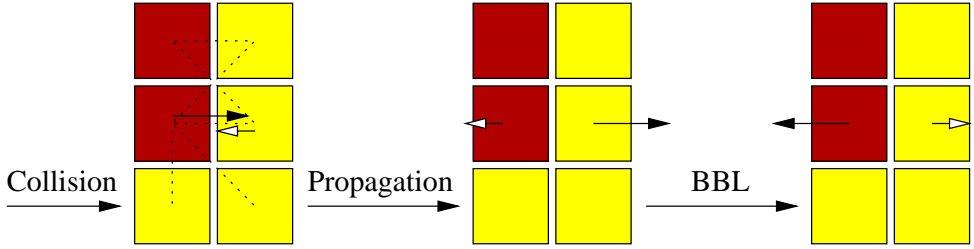}
\end{center}
\caption{The bounce-back on the links (BBL) algorithm: the dark sites
represent solid sites, the light sites are fluid sites; the dotted
lines on the leftmost picture outlines the location of the dry links.}
\label{fig:bbl}
\end{figure}

All solid objects are implemented applying stick boundary conditions,
following the bounce-back on the links (BBL) scheme proposed by
Ladd~\cite{Ladd94}. During propagation, the component of the
distribution function that would propagate into the solid node is {\em
bounced back} and ends up back at the fluid node, pointing in the
opposite direction. This produces stick boundary conditions at roughly
one half the distance along the link vector joining the solid and
fluid nodes, ensuring that the velocity of the fluid in contact with
the solid equals the velocity of the latter (see
figure~\ref{fig:bbl}). Let us assume that a solid-fluid boundary
exists between a node at $\vec{r}$ and one at $\vec{r} + \vec{c}_i$,
where $i$ labels the relevant lattice vector. Let $i'$ be the opposite
lattice vector, so that $\vec{c}_i = -\vec{c}_{i'}$. Then, at the
link, there are two incoming velocity distributions after the
collision; denote this time $t^+$. The post-collision distributions
are: $f_i (\vec{r},t^+)$ and $f_{i'}(\vec{r}+\vec{c}_i,t^+)$. These
distributions are \lq reflected' so that:
\begin{eqnarray}
\label{eqn:bbl}
f_i(\vec{r}+\vec{c}_i,t + 1) &=& f_{i'}(\vec{r}+\vec{c}_i,t^+)\\
f_{i'}(\vec{r},t + 1) &=& f_i(\vec{r},t^+) \nonumber
\end{eqnarray}

If the solid is moving with a velocity $\vec{u}_b$, the previous
boundary conditions have to be modified. If the densities $f_i$ and
order parameters $g_i$ are allowed to \lq leak' across the boundary
links, then the velocity at the link can be matched to the velocity of
the wall~\cite{Ladd94}.  In the case of a binary mixture, generalizing
the results of Ladd~\cite{Ladd94}, the basic BBL scheme is modified as
follows:
\begin{eqnarray}
\label{eqn:mod_bbl}
f_i(\vec{r}+\vec{c}_i,t + 1) &=& f_{i'}(\vec{r}+\vec{c}_i,t^+)
+ 6t^i \rho\vec{u}_b \cdot \vec{c}_i\\
f_{i'}(\vec{r},t + 1) &=& f_i(\vec{r},t^+)
- 6t^i \rho\vec{u}_b \cdot \vec{c}_i \nonumber\\
g_i(\vec{r}+\vec{c}_i,t + 1) &=& g_{i'}(\vec{r}+\vec{c}_i,t^+)
+ 6t^i \phi\vec{u}_b \cdot \vec{c}_i \nonumber\\
g_{i'}(\vec{r},t + 1) &=& g_i(\vec{r},t^+)
- 6t^i \phi\vec{u}_b \cdot \vec{c}_i \nonumber
\end{eqnarray}
\noindent where the quantities $t^i$ are geometric factors related to
the weights of the different subsets of velocities $\vec{c}_i$, and
are fixed when imposing the appropriate equilibrium distribution
functions for $f_i$ and $g_i$. Note that the BBL rules given above
require careful implementation if they are properly to account for the
effect on the composition variable $\phi$ of motion in a direction
normal to the solid-fluid boundary. This boundary condition allows us
to have a solid wall moving in any direction in contact with the
fluid~\cite{Warren}.

The velocity of the solid particles can be fixed beforehand. In this
case, one can use such moving objects \eg\ to apply a shear flow
through parallel plates at the boundaries of a sample, or to study
aspects of colloid hydrodynamics such as the steady-state
sedimentation of an ordered array of colloidal spheres with a
prescribed distribution. Alternatively, if the velocities of the solid
particles are updated, one can for example simulate the dynamics of
colloidal suspensions~\cite{Ladd94}.

\subsection{Boundary conditions}
Although periodic boundary conditions are applied to the model by
default, these can be modified by explicitly adding solid surfaces at
the boundaries. In the previous subsection we have shown how to add
them ensuring stick boundary conditions. This is enough for a
mono-component simple fluid. However, for complex fluids it is also in
general necessary to specify the behavior of additional fields at
solid boundaries, whether these are at the edges of the system or
internal boundaries between fluid and solid phases.

For example, in the case of a binary mixture it may be desirable to
specify the wetting properties of a solid surface, \ie\ its preference
for one of the two components. In order to deal with these situations,
we have developed in {\em Ludwig} a more generic way of characterizing
a solid interface. We have considered three different kinds of sites
on the lattice: solid, fluid and boundary sites (the latter are fluid
sites with at least one neighboring solid site). Accordingly, the
links are then classified as {\em wet} or {\em dry links} depending on
whether they join fluid sites or solid to fluid sites,
respectively. Then, in order to implement the appropriate
thermodynamic boundary conditions at the solid-fluid interfaces (which
we discuss in detail in the next section), the values of $f_i$ and
$g_i$ corresponding to the boundary sites and dry links are stored in
two separate lists, different from the basic vectors which store $f_i$
and $g_i$ for {\em all} sites. Note that additional information also
needs to be stored for the boundary links and boundary sites.  The
structure defined to store this information is called \keyw{bc\_link}.
In addition to the location and orientation of the links, it also
contains the force applied on the node and its velocity, in case it is
needed.
\begin{tabbing}
xxx \= xxx\= xxxxxxxxxx \= \kill
typedef struct bc\_link BC\_Link;\\

struct bc\_link\{\\
\> Int i,j,  \>\> /* i and j are the indices of the two nodes\\
\>           \>\> By convention, i is the (local) solid node,\\
\>           \>\> j is the fluid node */\\
\>  \> index,  \> /* index is the velocity that links the two nodes */\\     
\>    \> dup;  \> /* TRUE when a duplicate link */\\
\> Float frc,\>\> /* Force on the node at (r+0.5*c,t+0.5) in direction c */\\
\>       \> v; \> /* Velocity component of link */\\
\> BS\_Site *site; \>\> /* Only relevant to wetting: pointer to the site with the\\
\>          \>\> value of the concentration at the middle of the link */\\
\> BC\_Link *next; \>\> /* Store these in a singly linked list */\\
\};\\
\end{tabbing}

Note that all members of this list have a pointer \keyw{*site} to the
corresponding element of the boundary site list. Links that have their
ends on different PE domains (\ie\ partly in the halo region) need to
be duplicated on both PEs. The structure member \keyw{dup} is
therefore required to avoid multiple counting when carrying out
summations over all links. This structure is enough to implement the
BBL described in the previous subsection.

Similarly, boundary sites are stored in a structure of type
\keyw{bs\_site}, which includes information mostly required for the
implementation of the thermodynamic boundary conditions (wetting
effects). These include the free energy parameters of the neighboring
wall, the fluid or solid nature of the \keyw{NGRAD} neighboring sites
(\ie\ a binary map of dry or wet links to these sites), as well as the
actual gradient on the link which value is required by the BBL
algorithm described in section~\ref{ssec:solidobj}. The extended set
defined by \keyw{NGRAD} includes all vectors to nearest neighbors,
next nearest neighbors, next-but-one nearest neighbors, and the null
vector (thus for a 3-D cubic lattice (D3Q15), \keyw{NGRAD} = 27). This
extended set of neighbors has been introduced to improve the
representation of the order-parameter gradients close to the wall.
\begin{tabbing}
xxx \= xxxxxxxxxxxxxxxx \= \kill
typedef struct bs\_site BS\_Site;\\

struct bs\_site\{\\
\> Int ind;         \> /* (Global) index of site */\\
\> UInt wet;        \> /* Binary number specifying `wet' and `dry' links */\\
\> Float C,H;       \> /* Value of C and H in surface free energy */\\
\> Float phi[NGRAD];\> /* Gradients on the link (along NGRAD vectors) */\\
\>  BS\_Site *next; \>   /* Store these in a singly linked list */\\
\};\\
\end{tabbing}
Note that the actual implementation of the BBL is undertaken by
applying equation~\ref{eqn:mod_bbl} to all the (dry) links in the
linked list \keyw{BC\_Link}. Whilst the $f_i$ and $g_i$ can be
accessed directly using their array index, the link velocities
$\vec{u}_b$ and composition variable $\phi$ need to be retrieved from
the structure itself as \keyw{BC\_Link$\rightarrow$v} and
\keyw{BC\_Link$\rightarrow$site$\rightarrow$phi[]} respectively.

Another aspect worth pointing out is that both structures are defined
in different directories (see figure~\ref{fig:structure}). Indeed,
although the {\em boundary\_sites} functions are intrinsically
model-specific because they depend on the velocity sets and free
energy parameters, the {\em boundary\_links} routines on the other
hand are completely independent from the model used.

\subsection{Moving particles}
\label{ssec:movingobj}
The implementation of static solid objects and moving walls in
parallel is simple. Indeed, since these objects do not move across PE
domains during the course of a simulation, one needs only to build the
two linked lists \keyw{BC\_Link} and \keyw{BS\_Site} once, and
independently for each node.

The case of moving solid particles is more complicated and deserves
further discussion. Although non-spherical objects can be
described~\cite{vanderhoef}, we restrict the discussion to spherical
ones for simplicity's sake. In this case they are defined by four
parameter; their radius $R$, the position of their center of mass
${\bf r}_{CM}$ and their linear and angular velocities, ${\bf u}$ and
${\bf \omega}$.  Note that although ${\bf r}_{CM}$ moves continuously,
the surface of the particle is discretized since it is defined by the
lattice links which would cross the surface. Each particle can
therefore be defined as follows:
\begin{tabbing}
xxx \= xxx\= xxxxxxxxxxx \= \kill
typedef struct colloid Colloid;\\

struct colloid\{\\
\> Int index;\>\> /* Global index of colloid (0..N\_colloid-1) */ \\
\> FVector r,\>\> /* Position of the center of mass */\\
\>       \> v, w,\> /* Linear and angular velocity of the center of mass */\\
\>            \> T, F;\> /* Torque and force */ \\     
\> Float R; \>\> /* Radius */\\
\> Int \> *pe;\> /* List of PE domains span by this colloid */\\
\> Int \> local;\> /* TRUE for local PE, FALSE otherwise */\\
\> COLL\_Link *lnk;\>\> /* Pointer to list of links (surface) */\\
\> Colloid *next;\>\> /* Store these in a singly linked list */\\
\};\\
\end{tabbing}

\noindent where the list of links required to describe the surface of
each colloid is stored in a separate linked list of type
\keyw{coll\_link}:
\begin{tabbing}
xxx \= xxx\= xxxxxxxxxxxx \= \kill
typedef struct coll\_link COLL\_Link;\\

struct coll\_link\{\\
\> Int i,j,   \>\> /* i and j are the indices of the two nodes\\
\>            \>\> By convention, i is the inner node,\\
\>            \>\> j is the outer node */\\
\>  \> index,   \> /* index is the velocity that links the two nodes */\\  
\>  \> dup,     \> /* TRUE when a duplicate link\\
\>            \>\> (set when crossing different PE domains) */\\
\>  \> local;     \> /* TRUE for local PE, FALSE otherwise */\\
\> Float dist;\>\> /* Distance from middle of link to center of mass */\\
\> COLL\_Link *next; \>\> /* Store these in a singly linked list */\\
\};\\
\end{tabbing}

In order to update the velocities of the particle at each time step it
is necessary to know the force and torque exerted by the fluid on the
particle.  These are computed from the change in momentum of the fluid
at each link after it has bounced back. The motion of the colloid
evolves as in a molecular dynamics step. The discretization of the
particle surface onto the lattice links implies its radius is not
strictly constant. It is therefore necessary to keep the actual
distance between the center of the link and the center of mass of the
particle for {\em each} link as a separate variable \keyw{dist} to
compute the torque.

The complete information about any given colloid is replicated on all
the PEs which have their local domain intersected by it (list stored
in \keyw{colloid.pe}). The variable \keyw{coll\_link.local} tells
which (non-local) links must be skipped during the BBL, which is
implemented by applying equation~\ref{eqn:mod_bbl} just like for any
other solid objects.  Contributions to the torque from each link will
be summed locally and then across PEs.

The radius of the colloids must be larger than the lattice spacing,
and its actual value will depend on the volume fractions of solid
particles considered. Up to around $40\%$ in volume fraction, a small
value of the radius, \eg\ 2.5 lattice spacings, is sufficient to get
accurate results~\cite{hagen}. At higher volume fractions, larger
radii should be considered. This will eventually, at very high packing
fractions, limit the performance of the LB code, as progressively
larger system sizes will be needed. Similar limitations will apply
when dealing with polydisperse suspensions, or motion of non-spherical
objects.  A more complete discussion of the implementation,
optimization, and application of moving solid particles will be
published elsewhere.

\subsection{Wetting}
\label{ssec:wetting}
As pointed out already, for a two component fluid, the interaction
with a solid wall should allow a difference in interaction between the
two components and the wall even when the fluids are symmetric in all
other respects.  It can be energetically favorable for one of the two
components to be in contact with the solid surface, in which case, in
static equilibrium the fluid-fluid interface is not perpendicular to
the wall.  The equilibrium angle $\theta$ is called the contact angle
and is determined, via the Young equation~\cite{Rowlinson}
\begin{equation}
\label{eqn:young}
\gamma_{sl}^+-\gamma_{sl}^--\gamma_{ll}\cos(\theta) = 0
\end{equation}
\noindent where $\gamma_{sl}^{+(-)}$ is the solid/fluid surface
tension for the bulk phase with positive (negative) order parameter,
and $\gamma_{ll}$ is the fluid-fluid surface tension.

The resulting wetting phenomena are known to play a major part in the
behavior of complex fluids next to (or including) solid objects, but
their implementation in simulations still remains in its
infancy~\cite{Grubert,Yeomans95}. In particular, it is important to
make sure that the observed wetting behavior is consistent with the
thermodynamic requirement of Young's equation in equilibrium.  We have
therefore devised a novel predictor-corrector scheme to ensure an
accurate implementation of controlled wetting effects at the
solid-fluids interface in three dimensions.  Recalling that {\em
Ludwig} uses a symmetric $\phi^4$ model free energy (see
equation~\ref{eqn:fe}), a simple way to account for wetting properties
is to associate with the solid surfaces an additional surface free
energy density $f_s(\phi_s)$, where $\phi_s$ is the value of the
compositional order parameter in contact with the wall. According to
Cahn theory~\cite{Cahn}, the equilibrium order parameter profile
corresponds to that which minimizes the free energy functional $F =
F_{bulk}[\phi] + \int f_s(\phi_s) dS$ where $F_{bulk}$ obeys
equation~\ref{eqn:fe} and the integral is over the solid surface.  The
two solid-fluid interfacial tensions are found by minimizing this
expression near a flat solid-fluid interface to find the equilibrium
free energy $F$, subtracting the contribution $\int F_{bulk}[\phi^*]\,
d\mathbf{r}$ of the same volume of bulk fluid, and dividing by the
interfacial area.  The functional minimization also gives the
composition profile near the wall, and the boundary condition
satisfied at the solid surface, which is
\begin{equation}
\label{eqn:grad}
\frac{d {f}_s}{d \phi_s} = \kappa \left| \nabla\phi\cdot{\vec n} \right|
\end{equation}
\noindent where $\vec n$ is normal to the wall.

In general, ${f}_s$ is a function of the local order parameter. The
classical work on wetting has shown that a functional relation of the
form
\begin{equation}
\label{eqn:sfe}
{f}_s = \frac{C}{2} \phi_s^2 + H \phi_s
\end{equation}
\noindent is enough to reproduce the various different wetting
scenarios~\cite{Cahn}. By tuning the parameters $C$ and $H$, we modify
the properties of the surface in a thermodynamically controlled
manner~\cite{Cahn}, so that the fluid-solid interfacial tensions can
be tuned at will. Since we are dealing with a symmetric mixture, if
$H=0$ the two phases will have neutral wetting and show a local
variation in composition near the wall ($\phi_s - \phi^*$) of the same
magnitude. Nonzero $H$ allows then for an asymmetry in the surface
value of the order parameter for the two coexisting phases and a
contact angle different from $90$ degrees.

\begin{figure}[h]
\begin{center}
\includegraphics{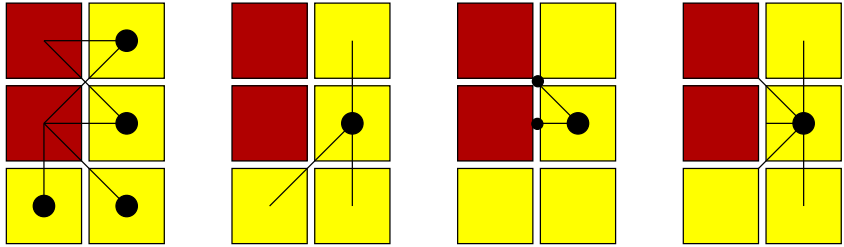}
\end{center}
\caption{The wetting algorithm.}
\label{fig:wetting}
\end{figure}

The main difficulty to implement the general boundary condition,
equation~\ref{eqn:sfe}, is that it depends on the value of the order
parameter at the surface, $\phi_s$, which is itself a dynamical
variable.  Moreover, due to BBL, the solid surface lies between the
sites thus making the calculation of $\nabla \phi$ and $\nabla^2 \phi$
by finite difference from neighboring sites using
equations~\ref{eq:grad_def} and~\ref{eq:lap_def} impossible. To
circumvent this, we use a predictor-corrector scheme to estimate the
gradient at the solid wall as follows (see figure~\ref{fig:wetting}):
\begin{enumerate}
\item determine which sites are next to a wall (boundary sites), and
hence which links cross the wall (\ie\, dry links);
\item estimate $\nabla \phi$ using finite differences on all wet
links;
\item from this estimate of $\nabla \phi$, extrapolate to halfway
along the dry links, and calculate $\phi_s$; using $\phi_s$ on the dry
links, calculate $\nabla\phi\cdot{\vec n}$ on these links;
\item calculate $\nabla \phi$ and $\nabla^2 \phi$ for the boundary
sites using {\em all} the gradients estimated on the links.
\end{enumerate} 

This scheme gives good quantitative results of the wetting angles in
accordance with thermodynamic predictions. Results from case studies,
both for a droplet and for planar interfaces are discussed in
section~\ref{sec:results}.

\subsection{Computational issues}
Production runs on the phase separation kinetics of binary fluid
mixtures have been carried out on the Cray T3D and the Hitachi SR-2201
at EPCC, and on the Cray T3E-1200 at
CSAR~\cite{Kendon00,Kendon99,Cates99}.  In addition to these
distributed memory systems, we also investigated the performance of
the code on shared-memory platforms such as the SUN HPC-3500 at EPCC.

\begin{figure}[h]
\begin{center}
\includegraphics[scale=0.8]{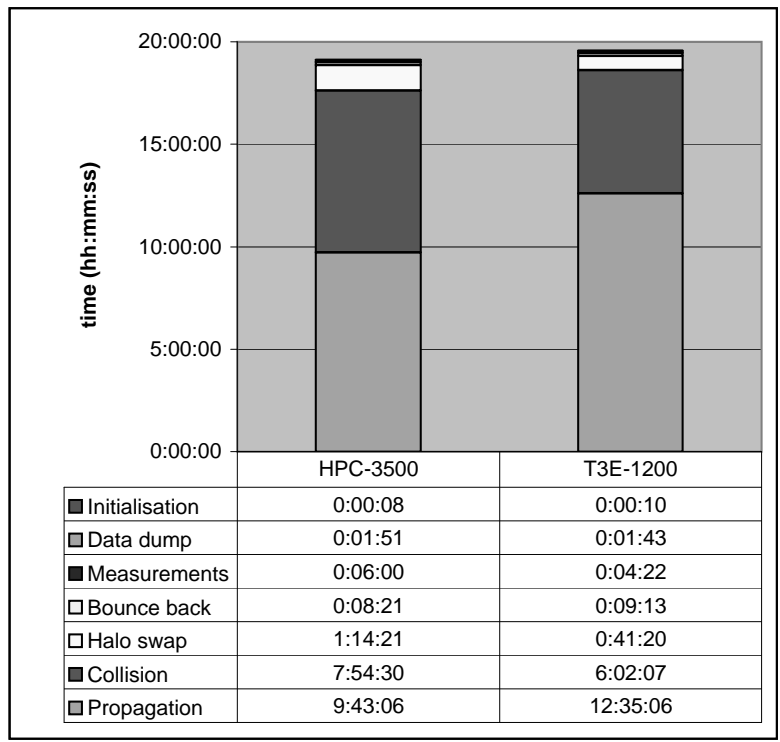}
\end{center}
\caption{Time spent in each function (average clock time per
processor). These timings have been obtained for the simulation of
spinodal decomposition under shear. The system was specified as
follows: D3Q15 model on a cubic lattice of size $L=180$ distributed on
8 PEs for 12,000 steps. The shear has been applied through fixed walls
at constant velocity, thus requiring a total of $2.7 \times 10^6$ BBLs
per step.  The order parameter and velocity fields were saved 60 times
and the full configuration only once, thus representing a combined
output of 17Gb for the simulation.  The darkest color at the top of
the bar chart represents the remaining stages of the simulation (\ie\
initialization, BBL, measurements, and I/O).}
\label{fig:benchmark}
\end{figure}

The profile information reproduced in figure~\ref{fig:benchmark}
provides us with an interesting insight on the critical sections of
the code. Firstly, one notices that over 90\% of the simulation time
is spent in the collision and propagation stages which are both
intrinsically serial and well load-balanced for most problems.  The
raw performance (excluding all I/O and measurements) varies from $1.04
\times 10^6$ to $1.10 \times 10^6$ gridpoint-updates per second for
the Cray T3E and SUN HPC-3500 respectively. Details of the timing
information for the various stages of the simulation of spinodal
decomposition under shear is given in figure~\ref{fig:benchmark}.
Note that the halo swap and BBL only account for 4-7\%. A comparison
of the profiles obtained on the Sun HPC-3500 and Cray T3E-1200 also
shows up that the critical parts of the code are highly system
dependent. As expected, the increased clock speed of the T3E-1200
(600MHz compared to only 400MHz on the HPC-3500) benefits the
collision stage which is a highly-localized algorithm with basic
arithmetic operators (add/multiply). This routine has been highly
optimized and makes a good use of the T3E memory hierarchy. On the
other hand, the memory-to-memory copies performed in the propagation
stage do not benefit from this increase in clock speed as much.
Indeed, the HPC-3500 significantly outperforms its rival by over 23\%
even though the algorithm for the propagation had been tuned for the
T3E by rearranging loops to make an efficient use of its streams
(see~\cite{CRI-streams} for further information about stream
optimization).  Particular attention should also be paid to finding
the optimal ordering for the velocity set $\{c_i\}$. It is important
to order the velocity set such that they correspond, as much as
possible, to sequential positions in memory for the distribution
functions. The first production platform for this package was the Cray
T3D which, due to its lack of second-level cache, was particularly
sensitive to data locality. The arrangement reproduced in
table~\ref{tab:velset} proved to be the most effective to preserve
data locality with a performance increase of over 20\% on the T3D
(compared to an unoptimized sequence) for the combined collision and
propagation stages. Note that some orderings can speed-up one of these
stages alone and be detrimental to the second one. The gain in
performance resulting from data locality is not as significant on
systems with second-level cache though. The \lq best' ordering of the
velocity vectors is therefore often system-specific.

\begin{table}[h]
\begin{center}
\caption{\lq Optimal' velocity set for the Cray T3D}
\label{tab:velset}
\begin{tabular}{|c|c|c|}
c(0) =( 0, 0, 0) & c(1) =( 1,-1,-1) & c(2) =( 1,-1, 1) \\
c(3) =( 1, 1,-1) & c(4) =( 1, 1, 1) & c(5) =( 0, 1, 0) \\
c(6) =( 1, 0, 0) & c(7) =( 0, 0, 1) & c(8) =(-1, 0, 0) \\ 
c(9) =( 0,-1, 0) & c(10)=( 0, 0,-1) & c(11)=(-1,-1,-1) \\  
c(12)=(-1,-1, 1) & c(13)=(-1, 1,-1) & c(14)=(-1, 1, 1) \\  
\end{tabular}
\end{center}
\end{table}

\begin{figure}[h]
\begin{center}
\includegraphics{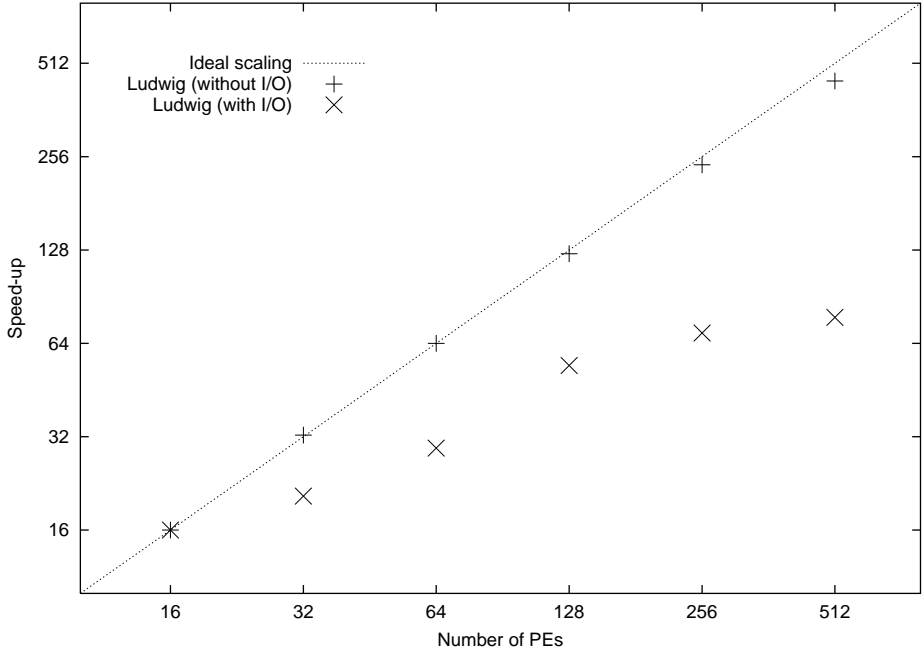}
\end{center}
\caption{Scale-up: scaling figures have been normalized with respect
to the time taken for the lowest number of processors,
$N_{PE}=16$. Details of the simulation are similar to that given in
figure~\ref{fig:benchmark} with the exception of the system size which
has been increased to $L=232$. The only quantity measured during this
benchmark is the order parameter field, $\phi_i$ which was measured
and saved to disk every 50 iterations. Speed-up data have been
included for simulations with and without the I/O (note that in both
cases, the time taken for dumping full re-start configurations was not
included). The I/O consists in 240 data dumps of 191 Mb written to
disks through eight parallel output streams.}
\label{fig:scaleup}
\end{figure}

As shown in figure~\ref{fig:scaleup}, {\em Ludwig} also demonstrates
near-linear scaling from 16 up to 512 processors. However, the overall
cost of the I/O can become a major bottleneck for the unwary (\eg\ a
$512^3$ system will generate in excess of 31Gb per configuration
dump).  The I/O has been optimized by performing parallel I/O. The
pool of PEs is split into $N$ groups of $p$ processors, thus providing
$N$ concurrent I/O streams (typically, $N=8$). Each group has a root
PE which will perform all I/O operations. The remaining $(p-1)$ PEs
send their data in turn to the I/O PEs which pack these data and write
them to disk. This approach usually offers high bandwidth without
having to use platform specific calls such as disk striping.

Note that MPI2-IO had initially to be discounted on the ground of
portability and performance.  We conclude this discussion by deploring
the lack of a full support for MPI-2 single-sided communications on
some platforms. Indeed, this functionality proves invaluable for the
implementation of the Lees-Edwards boundary conditions and moving
particles.


\section{Results for wetting behavior}
\label{sec:results}

{\em Ludwig} has already been used to study a number of problems for
binary fluid mixtures of current interest:
\begin{enumerate}
\item establishment of the role of inertia in late-stage
coarsening~\cite{Kendon00,Kendon99};
\item study of the effect of an applied shear flow on the coarsening
process~\cite{Cates99};
\item persistence exponents in a symmetric binary fluid
mixture~\cite{Kendon00b}.
\end{enumerate}
These results have been published elsewhere and will not be discussed
further here. A number of validation exercises relating to binary
mixtures are also described in~\cite{Kendon00}. Here we focus on
discussing new validation results obtained using the novel
predictor-corrector scheme for the thermodynamically consistent
simulation of wetting phenomena, as presented earlier.

We present results for two types of tests for the properties of binary
fluids near a solid wall. First, we verify that the modified
bounce-back procedure, equation~\ref{eqn:mod_bbl}, gives the expected
balance both for the momentum and for the order parameter. Second, we
check the numerical accuracy of the modified boundary condition that
accounts for the wetting properties of the wall,
equation~\ref{eqn:sfe}, as implemented through the predictor-corrector
step.

To test the validity of the modified bounce-back rule for the order
parameter field, we have looked at the motion of a pair of planar
interfaces perpendicular to two parallel planar solid walls when the
whole system is moving at a constant velocity parallel to the
walls. The initial condition corresponds to a stripe of one
equilibrium fluid ($\phi = \phi^*$), oriented perpendicular to the
walls, surrounded by a region of the other fluid ($\phi = -\phi^*$),
with periodic boundary conditions.  Due to Galilean invariance, the
profile should remain undistorted and move at the same velocity as it
is moving with initially, but with LB this is not guaranteed {\em a
priori} and has to be validated. We have considered both the case of
neutral wetting and the asymmetric case where $H$ is nonzero.

\begin{figure}[h]
\begin{center}
\includegraphics[scale=0.8]{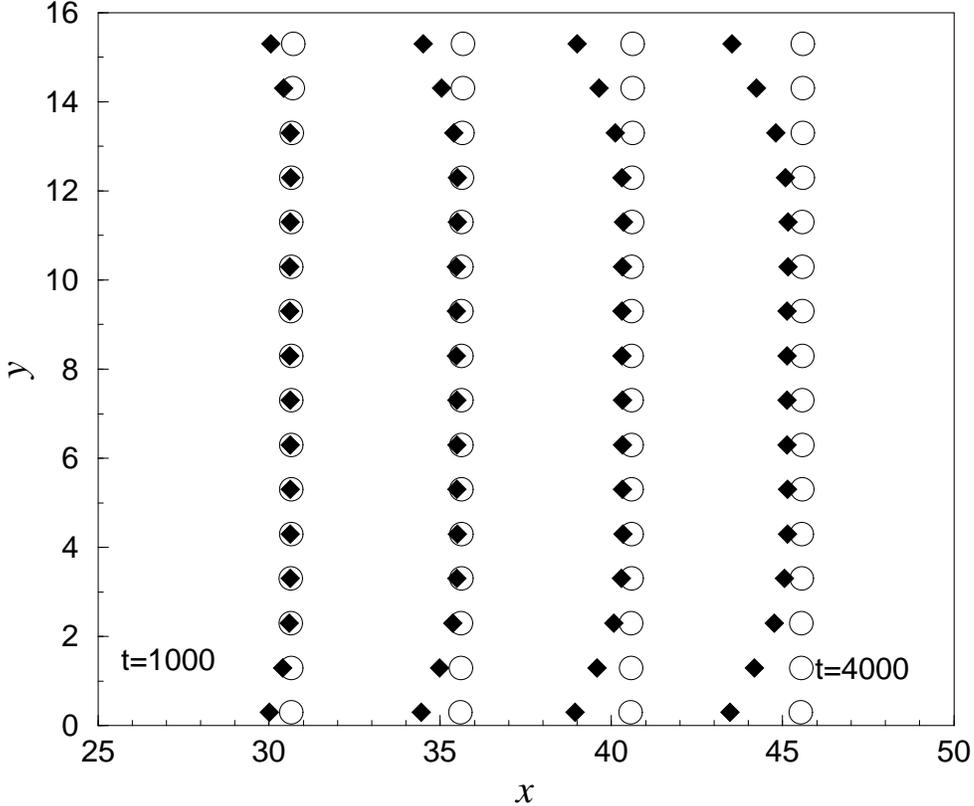}
\end{center}
\caption{Leading interface of a binary mixture fluid. Initially a
stripe perpendicular to two parallel moving walls is set up. The two
walls move at the same speed, 0.005 in lattice units, towards the
right. The two interfaces that characterize the stripes are recorded
every 1000 time steps (which corresponds to a theoretical displacement
of the two interfaces of 5 lattice units), and the leading one is
shown. The wall is neutral, which implies that no phase is favored at
the wall, and therefore the stripe wants to remain perpendicular. We
show the interfaces for the bounce-back described in
equation~\ref{eqn:bbl} (open circles), and also for a bounce-back of
the order parameter field in which the advection induced by the moving
walls is neglected (filled diamonds).}
\label{fig:interface1}
\end{figure}

For neutral surfaces, we show in figure~\ref{fig:interface1} the
leading interfacial profiles at different times, both for the
bounce-back rules described in equation~\ref{eqn:bbl}, and for an
improperly formulated bounce-back of the order parameter that does not
take into account the motion of the wall. (The latter is equivalent to
assuming $t^p=0$ only in the equations describing the bounce-back of
the order parameter distribution function.) As can be seen, although
the profile is advected due to the existence of a net momentum at the
wall, if the bounce-back is not done in the frame of reference of the
moving wall (leading to equation~\ref{eqn:mod_bbl}) the fluid-fluid
interface has a spurious curvature and it moves more slowly than it
should. From the rectilinear shape of both the leading (shown in
figure~\ref{fig:interface1}) and trailing interfaces of the
rectangular strip, we have found that Galilean invariance is in fact
well satisfied with
equation~\ref{eqn:mod_bbl}. Figure~\ref{fig:interface1} corresponds to
a situation of high viscosity and low diffusion ($\eta = 10, \tilde M
= 0.01$) but the same features have been observed for a number of
different physical parameters. The magnitudes of the errors made by
using the inappropriate bounce-back, in this geometry, are found to
decrease upon increasing the mobility, probably because a large
mobility allows a faster relaxation to the imposed velocity profile in
the interfacial region (especially near the contact with the walls).

\begin{figure}[h]
\begin{center}
\includegraphics[scale=0.8]{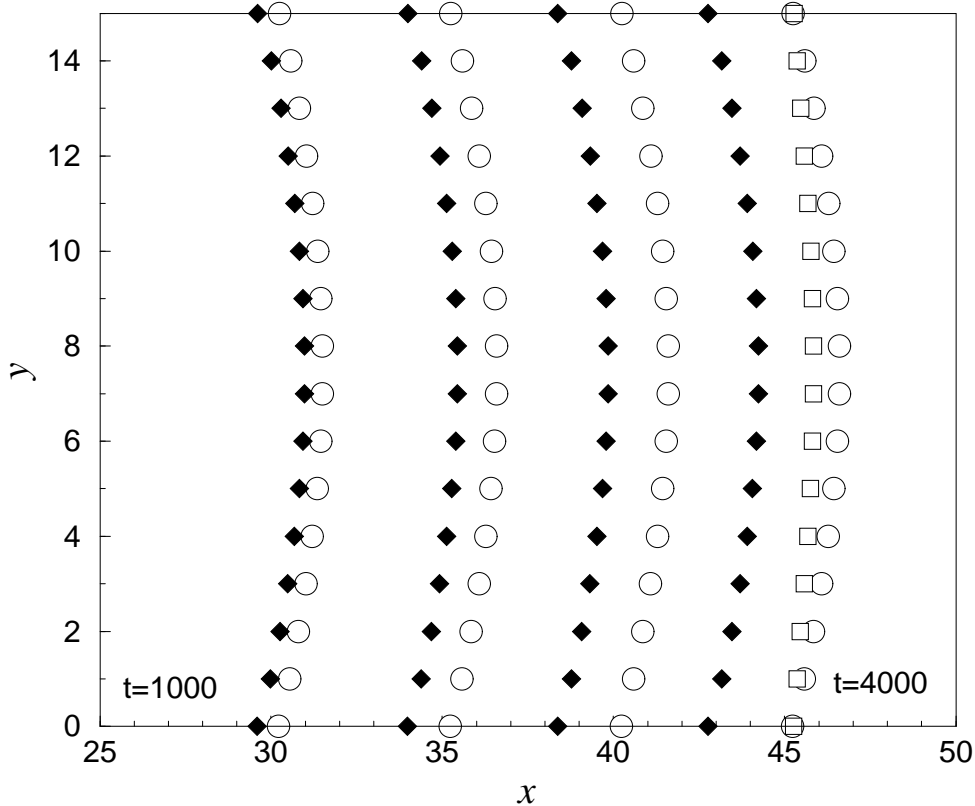}
\end{center}
\caption{Leading interface of a moving binary mixture fluid
stripe. Same geometry and parameters as in
figure~\ref{fig:interface1}.  The walls wet partially, and therefore
the stripe relaxes from its initial configuration to a bent
interface. We show the interfaces for the bounce-back described in
equation~\ref{eqn:bbl} (open circles), and also for a bounce-back of
the order parameter field in which the advection induced by the moving
walls is neglected (filled diamonds). In open squares we show the
trailing interface at time 4000 to show the magnitude of the
deviations with respect to Galilean invariance (see text).}
\label{fig:interface2}
\end{figure}

In figure~\ref{fig:interface2} we show the same configuration as
described in the previous paragraph, but for the case in which the
solid surface partially wets one of the two phases, so that $H$ is
nonzero and $\theta = 60 $ rather than 90 degrees. In this case the
profile should relax from the initial perpendicular stripe to a curved
interface. Again, the use of inappropriate bounce-back for the order
parameter leads to a slower motion of the interface, and a significant
distortion away from the equilibrium interfacial shape. In order to
test Galilean invariance here, for the final interface ($t=4000$
timesteps) we have compared the leading and trailing edge of the
stripe. There is a slight deviation in this case, implying that the
asymmetry entering via $H$ couples through to the overall fluid motion
relative to the underlying lattice, which (by Galilean invariance) it
should not. However, the resulting violation is very small, and the
interfacial deviations do not grow beyond one lattice spacing.

We have also verified that if the velocity of the walls is
perpendicular to their own plane, then an order parameter profile,
initially in equilibrium, remains stationary. This confirms that the
chosen boundary conditions can account for generically moving
interfaces for the case of a binary mixture.

\begin{figure}[h]
\begin{center}
\includegraphics[scale=0.8]{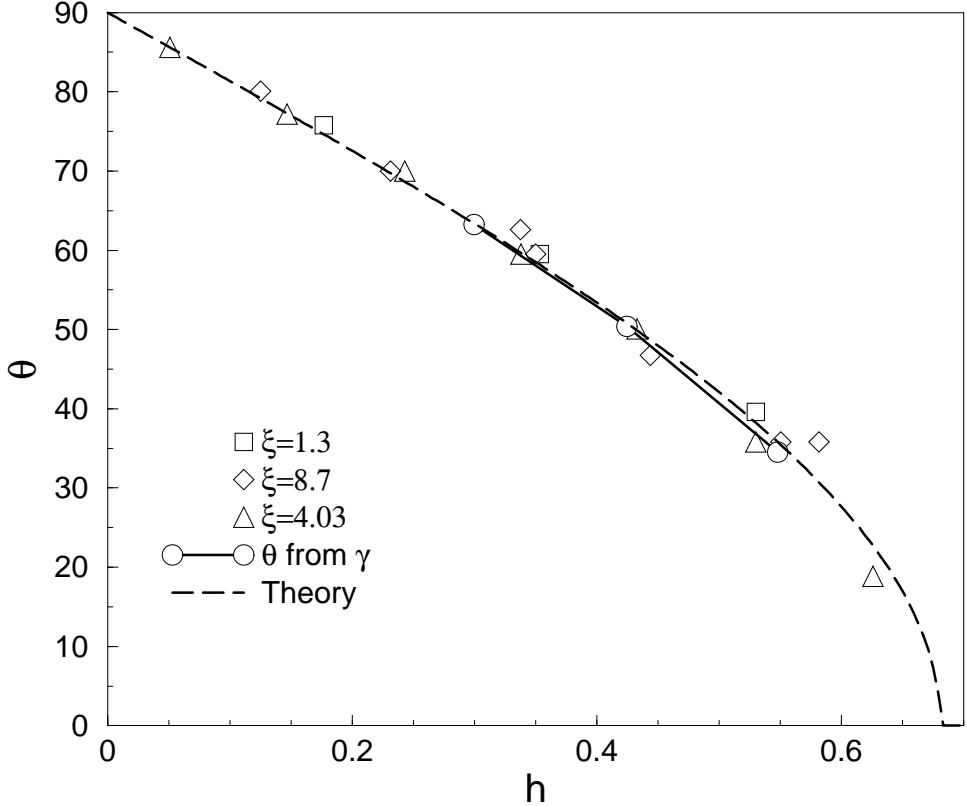}
\end{center}
\caption{Contact angles for the equilibrium profile of an interface in
contact with a wetting surface. $\xi=\sqrt{\kappa/B}$ is the
liquid/liquid interfacial width expressed in lattice units.}
\label{fig:angle}
\end{figure}

Finally, for stationary walls, we have computed the contact angles for
the simplest asymmetric case in which $C=0$ (see
equation~\ref{eqn:sfe}). In this situation the order parameter at the
wall will deviate by the same magnitude, but with opposite sign, in
the two bulk phases. For this choice (with $A = -B$) the contact
angle, $\theta$, is predicted to depend on the parameter
$h=H\sqrt{1/(\kappa B)}$ according to
\begin{equation}
cos(\theta) = \frac{1}{2}\left[-(1-h)^{3/2}+(1+h)^{3/2}\right] 
\end{equation}
We have considered a geometry in which the two solid walls have the
same wetting properties. We start, as in the previous case, with an
initial stripe perpendicular to the walls, defining two regions with
opposite equilibrium values for the order parameter. The equilibrium
profile for the interface then corresponds to a cylindrical cap. By
fitting the cylindrical cap it is possible to get a numerical value
for the contact angle. In figure~\ref{fig:angle} we show the measured
contact angles as a function of $h$ for different interfacial widths
$\xi_0 = (\kappa/2A)^{1/2}$, and compare these with the above
theoretical prediction. (Note that, by maintaining fixed $A/B$ have
kept constant the values $\pm \phi^*$ of the equilibrium order
parameters in the two coexisting phases.) As can be seen, the
agreement between the theoretical prediction and the measured contact
angle is quantitative.

The parameters that characterize the binary mixture have been chosen
to ensure fairly wide fluid-fluid interfaces. (In fact, the smallest
interfacial width, $\xi_0 = 1.3$, is at least twice as large as that
used previously in production runs for binary fluid
demixing~\cite{Kendon00}.)  For narrower interfaces than this, the
contact angles can differ significantly from the predictions due to
anisotropies induced by the lattice, whose effects were studied (for
fluid-fluid interfaces only) in~\cite{Kendon00}. This effect should be
more relevant for small contact angles. Indeed, when a narrow
fluid-fluid interface has a glancing incidence with the solid wall,
the discrete separation of lattice planes will lead to significant
errors in the estimation of the order parameter gradients; the
direction of these not only determines the surface normal of the
fluid-fluid interface, and hence the contact angle, but is crucial to
an accurate estimate of the free energies near the contact
line. However, except perhaps for small contact angles, there is not
much accuracy gained from choosing $\xi_0$ larger than $1.3$.

Because of the finite width of the interfaces, one has to be careful
to measure the contact angle by extrapolation from regions of
fluid-fluid interface that are more than about $\xi_0$ from the
wall. To check that we did this correctly, we have also numerically
and analytically computed the three interfacial tensions directly by
the method outlined in section \ref{ssec:wetting}.  From the values of
the surface tensions obtained, it is the possible to get values for
the contact angles, through Young's law, equation~\ref{eqn:young}.  As
expected, the values agree with the theoretical predictions and the
contact angles measured from the profiles (figure~\ref{fig:angle}).

\section{Conclusions}
We have described a versatile parallel lattice Boltzmann code for the
simulation of complex fluids.  The objective has been to develop a
piece of code that allows to study the hydrodynamics of a broad class
of multicomponent and complex fluids, focusing initially on binary
fluid mixtures with or without solid surfaces present.  It combines a
parallelization strategy, making it suitable to exploit the
capabilities of supercomputers, with a modular structure, which allows
its use without the need to know its computational details, and with
the possibility of focusing on the physical analysis of the
results. This strategy has led to a code that is in principle
adaptable to several different uses within the academic collaboration
involved~\cite{Kendon00,Kendon99,Cates99}.

We have discussed how to introduce generic fluid-solid boundary
conditions, and discussed which structures were developed to combine
the requirements of specific physical features with the generic
structure of the code.  The performance of the code in different
computers shows its portability, and it scales up efficiently on
parallel computers.

We have implemented generic boundary conditions for a binary mixture
in contact with moving solid interfaces. We have shown how one
recovers appropriate behavior of the momentum and the fluid order
parameter so long as the bounce-back rule, in the moving frame of the
wall, is performed with the distribution function that characterizes
the order parameter as well as that for momentum
(equation~\ref{eqn:mod_bbl}). A mesoscopic boundary condition that
accounts for the wetting properties of a binary mixture near a solid
surface has been described. It has been shown how to deal
appropriately with the gradients of the order parameter at the wall,
and with the role of the finite interfacial width when analyzing the
results. The values obtained for the contact angle agree with the
predictions of the model simulated, showing the absence of lattice
artifacts, at least for contact angles larger than about 20 degrees.
These results are, however, for planar solid interfaces oriented along
a lattice direction. We have not checked in detail the dependence of
the contact angle on the orientation of the solid surface, and it may
require further work on the discretization of order parameter
derivatives before this isotropy can be relied upon. Analogously to
the problem of small wetting angles mentioned above, the problem may
prove most acute for low-angle inclination of the solid surface, where
the naive discretization of the solid phase leads to a series of
well-separated steps in the wall position.

Current and planned work with {\em Ludwig} includes the hydrodynamic
simulation of multicomponent fluid flow in a porous networks with
controlled wetting; implementation of Lees-Edwards (sliding periodic)
boundary conditions; large-scale simulations of binary fluids under
shear, and the improvement of the gradients to make the thermodynamics
of this model more fully independent of the underlying symmetries of
the lattice. Longer term plans include studying colloid hydrodynamics
and extending {\em Ludwig} to study amphiphilic systems under shear
(see~\cite{Jury99b} for an example of this studied by DPD).

The authors would like to acknowledge Michael Cates, Simon Jury,
Alexander Wagner, Patrick Warren, and Julia Yeomans for valuable
discussions. They thank Michael Cates for assistance with the
manuscript. This work has been funded in part under the Maxwell
Institute's project on \lq Fluid Flow in Soft and Porous Matter' and
the EPSRC E7 Grand Challenge and GR/M56234.

\end{document}